\documentclass[%
reprint,
groupedaddress,
 amsmath,amssymb,
]{revtex4-2}

\usepackage{graphicx}
\usepackage{dcolumn}
\usepackage{bm}
\usepackage{hyperref}

\hypersetup{
    colorlinks=true,
    linkcolor=blue,
    filecolor=magenta,      
    urlcolor=cyan,
    pdftitle={Sharelatex Example},
    bookmarks=true,
    citecolor=blue
}

\begin{document}

\preprint{APS/123-QED}

\title{Ultimate Phase Sensitivity in Surface Plasmon Resonance Sensors by Tuning Critical Coupling with Phase Change Materials}

\author{Lotfi Berguiga$^1$}%
\email{lotfi.berguiga@insa-lyon.fr}
\author{Lydie Ferrier$^1$}%
\author{Cécile Jamois$^1$}%
\author{Taha Benyattou$^1$}%
\author{Xavier Letartre$^2$}%
\author{Sébastien Cueff$^2$}%

\affiliation{$^1$Université de Lyon, Institut des Nanotechnologies de Lyon - INL, CNRS UMR 5270, INSA Lyon - CNRS, 69621 Villeurbanne, France}%
\affiliation{$^2$Université de Lyon, Institut des Nanotechnologies de Lyon - INL, CNRS UMR 5270, Ecole Centrale de Lyon - CNRS, 69134 Ecully, France}

\date{\today}

\begin{abstract}
Plasmonic sensing is an established technology for real-time biomedical diagnostics and air-quality monitoring. While intensity and wavelength tracking are the most commonly used interrogation methods for Surface Plasmon Resonance (SPR), several works indicate the potential superiority of phase interrogation in detection sensitivity. Here, we theoretically and numerically establish the link between ultra-high sensitivities in phase interrogation SPR sensors and the critical coupling condition. However, reaching this condition requires a technically infeasible angstrom-level precision in the metal layer thickness. We propose a robust solution to overcome this limitation by coupling the SPR with a phase-change material (PCM) thin film. By exploiting the multilevel reconfigurable phase states of PCM, we theoretically demonstrate ultra-high phase sensitivities with a limit of detection as low as $10^{-10}$ refractive index unit (RIU). Such a PCM-assisted SPR sensor platform paves the way for unprecedented sensitivity sensors for the detection of trace amounts of low molecular weight species in biomedical sensing and environmental monitoring.
\end{abstract}


\maketitle

\section{Introduction}
Detection and measurement of chemical species is of paramount importance in diverse areas such as biosensing for cancer monitoring \cite{Bellassai2019} or for the environmental detection of volatile organic compounds (VOC) \cite{Brenet2018}. 
Owing to its large sensitivity to refractive index modifications, surface plasmon resonance (SPR) is one of the most widely exploited effect in optical sensors applications, especially for label-free detections of chemical and biological species. Among other features, SPR enables quantitative measurements of affinity and kinetics of biomolecular binding as for example in antibody-antigen, ligand-receptor kinetics or in enzyme-substrate reaction \cite{Homola2008}. As such, it has been applied in fundamental biological studies, drug screening, medical diagnostics, environmental monitoring, and food safety and security  \cite{Homola2008, Singh2016, Bellassai2019}. 

However, even though impressive performances were achieved with SPR sensors, improving sensitivity remains an active field of research for this technology -- and for optical sensors in general. Accurately measuring low molecular weight compounds or small quantities of chemical species would lead to breakthrough developments in early stage cancer diagnostics or in early detection of hazardous chemicals in air for environmental monitoring and health security. \\

While intensity and wavelength tracking are the most commonly used interrogation methods, many experimental works indicate the potential superiority of SPR phase interrogation in detection sensitivity. Interferometric and polarimetric methods have been developed to measure the optical phase of SPR reflected light \cite{Kabashin1998,Kabashin1999,Nikitin1999,Ho2002,Wu2003,Wu2004,Lee2007,Markowicz2007,Ng2013}.
In phase interrogation SPR, the average limit of detection (LOD) is around $10^{-8}$ refractive index unit (RIU) \cite{Zeng2017} and a few very low LOD -- such as $2.8\times10^{-9}$ RIU by Li et al. \cite{Li2008} -- were reported. Despite these low experimental LOD for phase interrogation, up to now no theoretical studies demonstrate that this method provides a quantitatively larger sensitivity than direct intensity interrogation (wavelength or angular SPR spectroscopy) although phase and intensity interrogation have been compared \cite{Ran2006,Nikitin1999,Kabashin2009}. 

In the following, we derive an analytical framework that explicitly establishes a direct relation between the SPR phase sensitivity and the SPR minimum of reflectivity at resonance as well as a relation between phase sensitivity and the usual figure of merit (FOM) for optical sensors. This formal link reveals that phase interrogation should lead to unprecedented sensitivity levels for the measurement of refractive index modifications (down to $10^{-10}$ RIU). However, reaching this metric would require controlling the metal thickness to a sub-nanometric precision, which is technically unrealistic. Here, we propose an original hybrid SPR platform in which ideal critical coupling conditions can actively be achieved by a direct reconfiguration technique using phase-change materials (PCM) thin films.

\begin{figure}[htbp]
\centering\includegraphics[width=8cm]{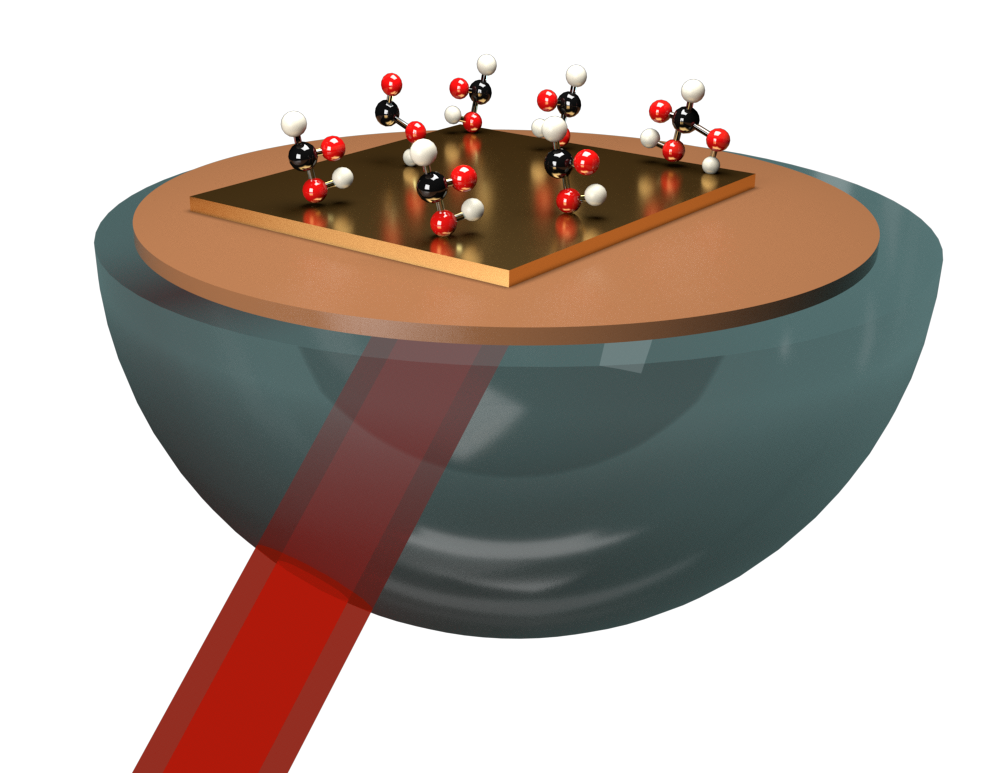}
\caption{a) Surface plasmon resonance excitation scheme in the Kretschmann configuration for a PCM-assisted SPR sensor system with a thin GST layer sandwiched between the coupling glass medium and the gold layer}
\label{fig:SPRdescription}
\end{figure}

Phase change materials provide interesting pathways for reconfigurable light management in nanophtonic systems such as active beam-steerers ~\cite{de2018nonvolatile}\cite{abdollahramezani2020dynamic}, bi-directional all-optical switching ~\cite{gholipour2013all}, dynamic modulation of light emission ~\cite{cueff2015dynamic}, light absorption~\cite{carrillo2018reconfigurable,john2020multipolar} or light transmission~\cite{howes2020optical}. Guo et al. opened the road to PCM-based sensors by numerically showing  the shift of a gap surface plasmon resonance of a metamaterial-based perfect absorber in the mid-infrared range using standard reflection intensity measurement \cite{Guo2018}. Sreekanth et al. \cite{Sreekanth2018b} experimentally measured phase jumps of reflectivity when switching GeSbTe (GST) from amorphous to crystalline. Moreover they measured the Goos-Hänchen shift, which is directly related to the derivative of the optical phase \cite{Sreekanth2021}. To go beyond simple binary states of PCM, it has been shown that nonvolatile multilevel phase states of GST enables extremely precise tuning to reach deep perfect absorption in lithography-free devices \cite{Cueff2020}. 
In the present work, we propose a concept leveraging a new SPR-based PCM plateform for phase interrogation, as sketched in Fig.\ref{fig:SPRdescription}. Combining the phase interrogation SPR approach with reconfigurable phase sensitivity by post-tuning the critical coupling of SPR via multilevel GST phase states paves the way for SPR optical sensors with unprecedented sensitivities. 

In a first part, we revisit the theory of SPR reflectivity in angular interrogation and establish an analytical relation between phase sensitivity and minimum of reflectivity. This formalism identifies the crucial role of the critical coupling condition to reach ultra-high sensitivities.
In a second part we describe the optical properties of GST and their evolution from amorphous to crystalline state as well as in several intermediate partially crystallized states. The phase sensitivity of the device is numerically analyzed for small refractive index variations and for various crystalline fractions of GST. The fine tuning of GST phase states allow us to design extremely sensitive hybrid GST-SPR sensors. We finally discuss the implications and future implementations of such PCM-assisted SPR sensors.

\section{SPR phase sensitivity}

A surface plasmon at the interface between a semi-infinite metal and a dielectric medium follows the dispersion relation:

\begin{equation}
n^{sp}_{eff}=\sqrt{\frac{\epsilon_m \epsilon_{2}}{\epsilon_m+\epsilon_{2}}}
\label{SPRconditions}
\end{equation}
With $\epsilon_{m}$ and $\epsilon_{2}$ the optical dielectric constants of the metal layer and the dielectric medium, respectively.

In the Kretschmann geometry, using the total internal reflection method, an incident light wave incoming with an angle of incidence $\theta$ from the coupling medium (usually a prism) with refractive index $n_{0}$ will excite a surface plasmon wave at the interface of a thin metal layer when $\theta$ equals the angle $\theta _p$ i.e. when the following coupling condition is fulfilled 

\begin{equation}
n_{o}sin (\theta_p)=n^{sp}_{eff}
\label{SPRconditions2}
\end{equation}

\begin{figure}[htbp]
\centering\includegraphics[width=\linewidth]{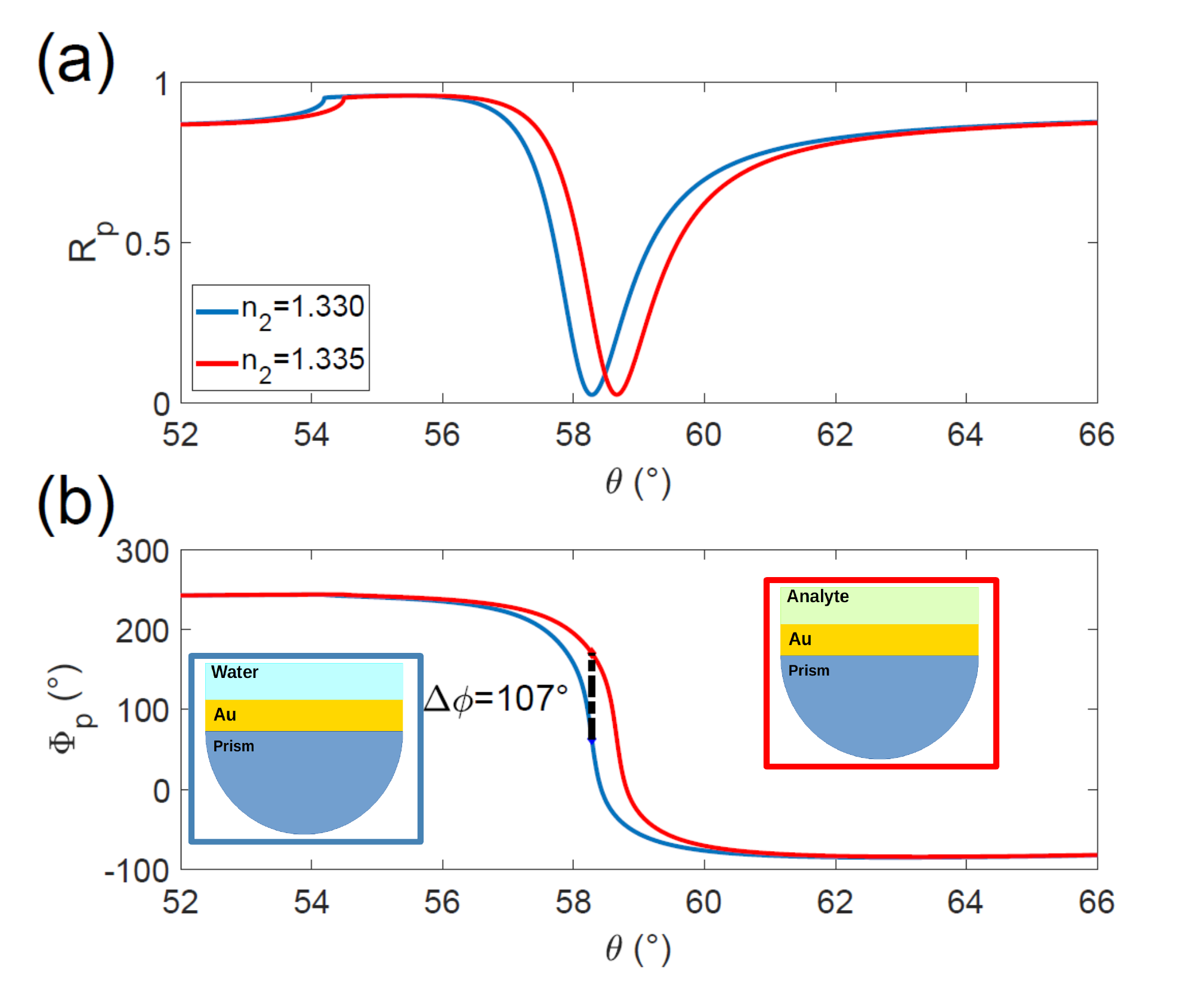}
\caption{Surface plasmon resonance shift versus refractive index variation of analyte media from $n_2=\sqrt{\epsilon_2}=1.33$ to 1.335  for a  47 nm gold layer on a prism of refractive index $n_0=1.64$. Reflected intensity in a) and  reflected phase in b). The wavelength is 750 nm}
\label{fig_SPRcurves}
\end{figure}

Consider a prism on which is deposited a  gold layer, which is directly in contact with an analyte (Fig. \ref{fig:SPRdescription}(a)).  At the SPR angle, where the light is coupled to a surface plasmon, the reflected light presents a dip in intensity and an abrupt phase jump, as shown in Fig.\ref{fig_SPRcurves}. The reflected light versus the the angle of illumination follows the Lorentzian approximation as described in  \cite{Raether1988,Homola2006}:


\begin{equation}
R(\theta)=1-\frac{4\gamma_{i}\gamma_{rad}}{(n_o sin\theta-n^{sp}_{eff})^2+(\gamma_{i}+\gamma_{rad})^2}
\label{SPR_ReflectvityEquation}
\end{equation}
With $\gamma_i$ and $\gamma_{rad}$, the leakages due respectively to surface plasmon absorption in the metal and radiative leakage (reconversion of surface plasmon in radiative light trough the prism).

Numerical simulations have been done with a home-made software based on the transfer matrix method \cite{Yeh2005} to calculate the complex reflection coefficients as a function of the incident angle, the wavelength and the polarization.\\

\begin{figure*}[htb]
\centering\includegraphics[width=16cm]{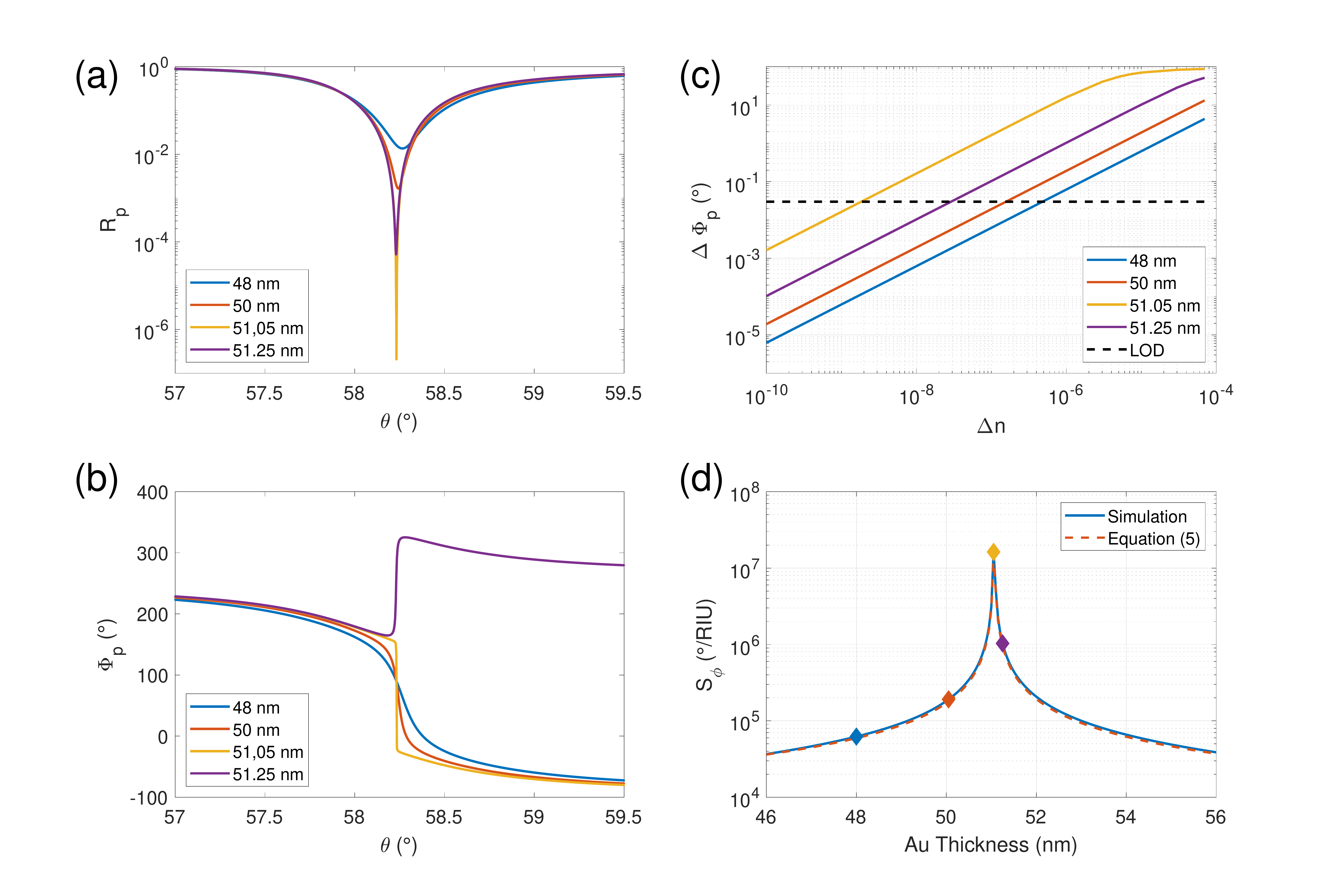}
\caption{Surface plasmon resonance and sensitivity as a function of gold thickness and refractive index of the analyte. a) Reflected intensity and b) phase of SPR for 4 different gold thicknesses. c) Phase variation as a function of refractive index variation of the analyte medium at initial SPR resonance angle $\theta _{p}$ for  the same 4 different gold thicknesses. The dashed curve is the limit of measurable phase value  $\Delta{\phi}=3\sigma_{\phi}=0.03^{\circ}$. d) Sensitivity evolution $S_{\phi}$ with gold thickness : numerical simulations (blue curve) and analytical model (dashed red curve). The diamond markers correspond to the sensitivity of the four curves in c).}
\label{fig_SPRsensitivity}
\end{figure*}

 SPR is very sensitive to changes at the metal/analyte interface: when the refractive index of the analyte is modified, the resonance condition changes and leads to an angle shift. For example in the case of an SPR supported by a 47-nm-thick gold layer, a variation of the refractive index of the analyte as small as 0.005 will lead to an angle shift of 0.38$^{\circ}$ (Fig. \ref{fig_SPRcurves}(a). This behavior is also retrieved in the phase of the reflected light, as shown in Fig. \ref{fig_SPRcurves} (b). The phase shift for the same refractive index variation of the analyte is $107^{\circ}$. 
 Intuitively, one finds that the sharper is the resonance, the higher is the phase variation upon changing the refractive index of the dielectric medium \cite{Nikitin1999}. \\
 The slope of the phase step at SPR resonance can be drastically enhanced by the critical coupling, a condition for which perfect absorption occurs when the metal absorption equals the radiative leakage ($\gamma_i=\gamma_{rad}$). For a given wavelength, the conditions for perfect absorption are met for a precise metal layer thickness. Perfect absorption in photonics is usually claimed once the reflectivity value approaches zero. However, to precisely gauge a very low level of reflection, one should go beyond this depiction and represent results in logarithmic scale.  To illustrate this, Fig. \ref{fig_SPRsensitivity}(a) shows the log-scale p-polarized reflection of thin gold layers with varying thickness in the Kretschmann configuration.  As displayed, very low reflectivity values are obtained for all gold thicknesses between 48 nm and 52 nm. However, the reflectivity values are strongly impacted by the gold thickness, and extremely low levels of reflectivity (lower than 10$^{-6}$) corresponding to a quasi-perfect absorption, are reached precisely for a 51.05 nm-thick gold layer. 
 These conditions of perfect absorption lead to a very sharp phase step at the resonance (see Fig. \ref{fig_SPRsensitivity}(b)). To quantify the SPR phase sensitivity $S_{\phi}$ versus the gold thickness, we evaluate the phase variation $\Delta \phi_{p}$ at the SPR angle when the refractive index of the analyte medium is modified. Using water as the analyte reference and applying a small modification $\Delta n$ to its refractive index, the corresponding phase variation reads as follows: $\Delta \phi_{p}=\Phi(n_2=n_{water}+\Delta n)-\Phi(n_2=n_{water})$ at $\theta_{p}$ of water. In Fig. \ref{fig_SPRsensitivity}(c) the phase variations as a function of $\Delta n$ are displayed for the four gold thicknesses previously shown in Fig. \ref{fig_SPRsensitivity}(a) and b). 
 We can notice in Fig. \ref{fig_SPRsensitivity}(c) that the phase shift is linear in a wide range of refractive index variation and strongly depends on the gold thickness. The phase shift reaches a maximal variation for the optimal gold thickness of 51.05 nm, which exactly corresponds to the critical coupling condition for SPR. Fig. \ref{fig_SPRsensitivity} (a), (b) and (c) clearly demonstrate that there exists an optimal gold thickness where the phase variation is the highest, which one corresponds to a reflectivity minimum, in accordance to the perfect absorption condition. 

 To quantify the phase shift with standard metrics, we define the bulk sensitivity $S_{\phi}$ as the slope of phase variation curves:
\begin{equation}
S_{\phi}=\frac{\Delta{\Phi}}{\Delta{n}}
\label{SensitivityDefinition}
\end{equation}
Fig. \ref{fig_SPRsensitivity}(d) shows the corresponding sensitivity $S_{\phi}$ as a function of gold thickness. 
As the phase variation drastically increases when SPR reaches the critical coupling, the sensitivity $S_{\phi}$ is raised by three orders of magnitude and culminates to values larger than  $1.5\times10^{7}$ $^{\circ}$/RIU .
 
 State of the art SPR setups enable the detection of a $\Delta{\phi}=0.03^{\circ}$ phase modulation (represented by a dashed black curve in Fig. \ref{fig_SPRsensitivity}(c) ). Values of $\Delta n$ producing such a phase modulation therefore correspond to the limit of detection (LOD) -- which represents the minimal refractive index change $\Delta n_{min}$ than can be detected.
 In the condition of critical coupling described previously (corresponding to the yellow curve in Fig. \ref{fig_SPRsensitivity}(b) ), we can therefore expect a LOD equal to  $10^{-9}$ RIU. The LOD is related to the bulk sensitivity of the sensor $S_{\phi}$ and to the noise of the experimental set-up $\sigma_{\phi}$: $LOD=\Delta n_{min}=3\sigma_{\phi}/{S_{\phi}}$.

Using coupled mode theory (CMT), we can model the system as a resonator in which the leakage rate due to surface plasmon absorption is characterized by $\gamma_i$ and the radiative leakage rate by $\gamma_{rad}$. From this model, we find an expression for the complex reflection coefficient (see Appendix A) and we also derive an original relation between the phase sensitivity and the minimum of reflectivity. To get this relation, we first express the derivative of the phase as a function of wavelength. After some algebraic manipulations in order to rely phase derivative to phase sensitivity, we extract an expression for the sensitivity at the resonance wavelength that is inversely proportional to the minimum of reflectivity (See Appendix B). The relation is :
\begin{equation}
S_{\phi} =\frac{2Q}{\lambda_{sp}}\frac{S_{\lambda}}{r_{min}}
\label{equSensitivityvsReflectivity}
\end{equation}
With $Q$ the quality factor, $S_{\lambda}$ the bulk sensitivity in wavelength interrogation ($S_{\lambda}=\frac{\Delta{\lambda}}{\Delta{n}}$), $\lambda_{sp}$ the SPR wavelength and $r_{min}$ the reflection coefficient at resonance ($r_{min}=\sqrt{R_{p}(\theta _{p})}$). The theoretical curve with a coefficient $7015~^{\circ}/RIU$ superposes very well with the numerical simulation (Fig. \ref{fig_SPRsensitivity}(d)). 
This formalism conveys two important informations for the first time to the best of our knowledge: i) a relation between phase sensitivity $S_{\phi}$ and wavelength sensitivity $S_{\lambda}$; ii) a direct quantitative link between the phase sensitivity and the minimum of reflectivity (even though it was pointed out empirically in previous works)

With equation \ref{equSensitivityvsReflectivity}, we understand in more details the role of the minimum of reflectivity: it acts as a gain factor on the phase sensitivity. Said otherwise, the phase sensitivity $S_{\phi}$ is the equivalent of the wavelength sensitivity $S_{\lambda}$ amplified by a gain factor $1/r_{min}$. As a direct consequence, phase sensitivity becomes much higher than wavelength sensitivity when $r_{min}$ converges to zero i.e. towards critical coupling. Moreover, this compact formal expression explicitly shows the role of the quality factor $Q$ in phase sensitivity. As we have established a direct relation between phase sensitivity and wavelength sensitivity,  we can reformulate the expression for the phase sensitivity :
\begin{equation}
S_{\phi} = 2\frac{FOM}{r_{min}}
\label{equSensitivityvsReflectivity2}
\end{equation}
with the widely used figure of Merit (FOM) defined as $FOM=Q\times S_\lambda/\lambda_{0}$.

Equation \ref{equSensitivityvsReflectivity2} implies that very sensitive SPR sensors in wavelength interrogation will lead to even more sensitive sensors in phase interrogation. \\

 The simulations results clearly indicate that reaching such an ultimate phase sensitivity requires adjusting the metal thickness at an angstrom-level precision (Fig. \ref{fig_SPRsensitivity}(d)). Specifically, a deviation of 0.2 nm between the targeted and fabricated thickness leads to a reduction in the phase sensitivity by more than one order of magnitude  (from $1.6\times10^7$ to $10^6$ $^{\circ}$/RIU). Such a level of precision in metal layer deposition is hardly achievable, not to mention that the surface roughness of metal thin films typically exceeds a few nanometers. For these two reasons, the possibility to produce fabrication-tolerant SPR phase sensors with ultimate sensitivity is unreachable. In an attempt to experimentally demonstrate SPR phase sensors with very high sensitivity, Huang et al. fabricated several samples with gold thickness varying by 0.3 nm steps. By selecting the most sensitive sample they achieved a sensitivity of $5.10^6$ $^{\circ}/RIU$ \cite{Huang2011}. Even for such a rigorous and systematic study, our simulations indicate that a bit of luck was necessary to demonstrate such a high sensitivity, given that a 0.3 nm step is not sufficiently accurate to target sensitivity over one million.
 
 This stringent thickness-dependence may explain the discrepancies in the measured LOD reported by different groups -- ranging from $5\times10^{-7}$ to $10^{-9}$ RIU (this discrepancy can be noticed in the review paper by  Zeng et al. \cite{Zeng2017} and references therein where a summary of LOD in RIU is collected in phase interrogation SPR). 
 \\
 
To summarize, the theory presented here shows that high sensitivities can be obtained using SPR sensors with a phase interrogation method by precisely adjusting the critical coupling via the gold layer thickness. However, demonstrating ultimate sensitivities appears technically infeasible due to the angstrom-level accuracy needed for the thickness of the metal layer. 
In the following, we show that introducing phase change materials in the device not only enables overcoming this experimental limitation but also considerably expand the fabrication tolerance and overall robustness of the device to reach ultimate sensitivity.

\section{Phase-Change-Materials-assisted SPR}

\subsection{Optical properties of GST}
\begin{figure}[htb]
\centering\includegraphics[width=7cm]{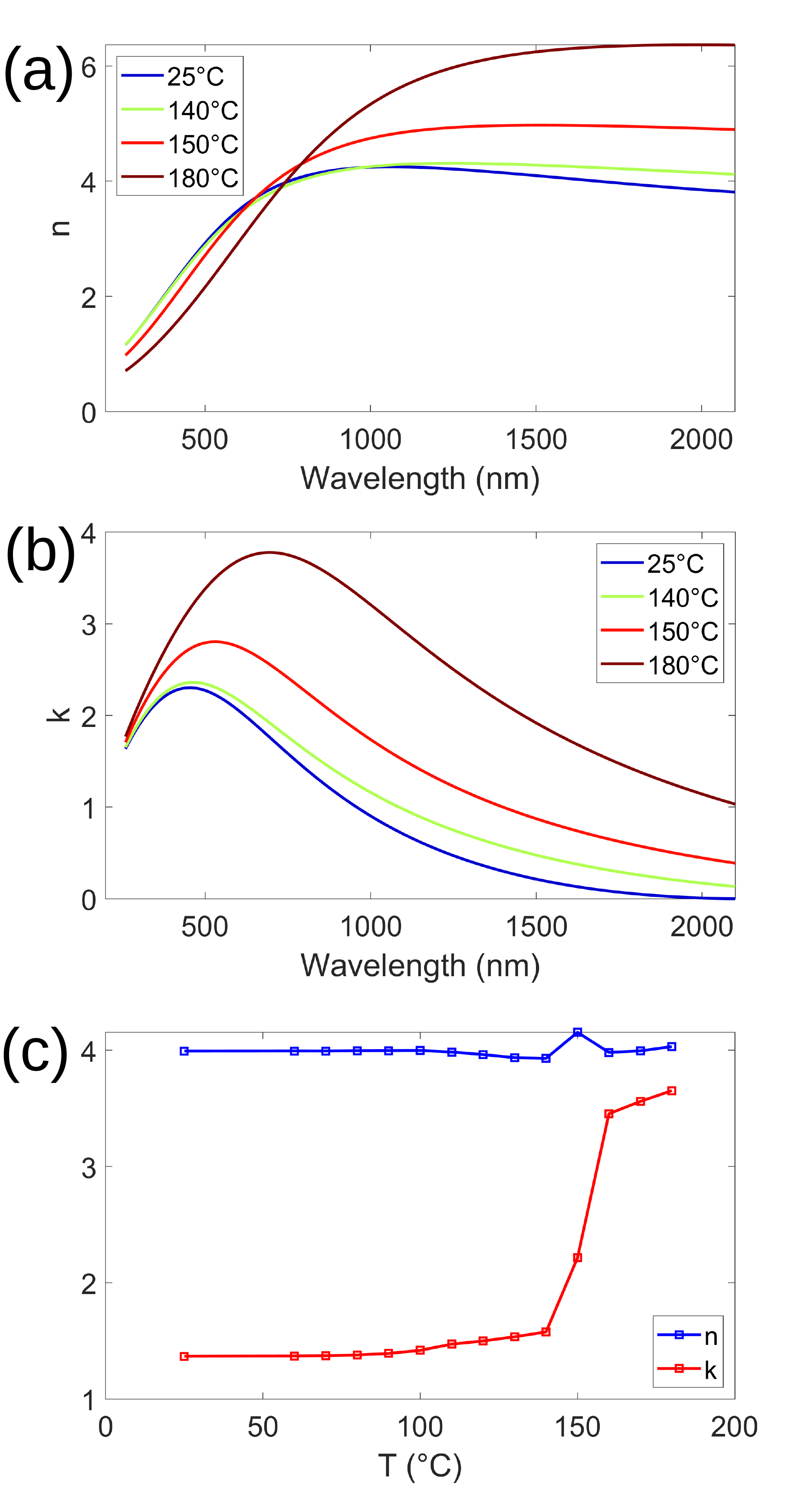}
\caption{Optical properties of GST thin film. Dispersion curves of n (a) and k (b) versus temperature. c) Evolution of n and k as a function of temperature at a wavelength of 750 nm .}
\label{fig_GSTdispersion}. 
\end{figure}

Phase change materials such as GST present specific chemical bonding properties in which the long range order between atoms produces a very large refractive index, typically around $n\sim$6-7 in the infrared region ~\cite{shportko2008resonant,huang2010bonding,zhu2018unique, raty2019quantum}. As this long range order is suppressed when the material is in an amorphous state, the refractive index is strongly reduced i.e. $n\sim$4. The crystallization-amorphization of GST therefore produces a very large modulation of the complex refractive index. Note that once the GST is crystallized (partially or entirely), it will stay in that state even when the sample is brought back to room temperature. This non-volatile nature of the tunable refractive index is an important feature that we will use in that study.
Fig. \ref{fig_GSTdispersion} (a) and (b) illustrate the drastic change of refractive index $n$ and absorption coefficient $k$ of a GST layer  upon crystallization, as measured by spectroscopic ellipsometry as a function of temperature \cite{Cueff2020}. 

Most of the works using GST for tunable photonics leverage the refractive index modulation and try to mitigate the optical absorption. In the present work, we do not use the refractive index modulation but rather exploit the associated increase in extinction coefficient as a means to adjust the optical absorption.
 
 To mimic numerically  the evolution of $n$ and $k$ as a function of the wavelength in the phase transition range we have reasonably approximated it  with a linear regression \cite{Cueff2020}. Using this linear variation $n$ vs $k$ we introduce the quantity $\tau$, that represents the crystallization fraction where 0 and 100 \% correspond to the amorphous and crystalline state, respectively.\\

\subsection{Tuning critical coupling with GST}

\begin{figure}[htb]
\centering\includegraphics[width=8cm]{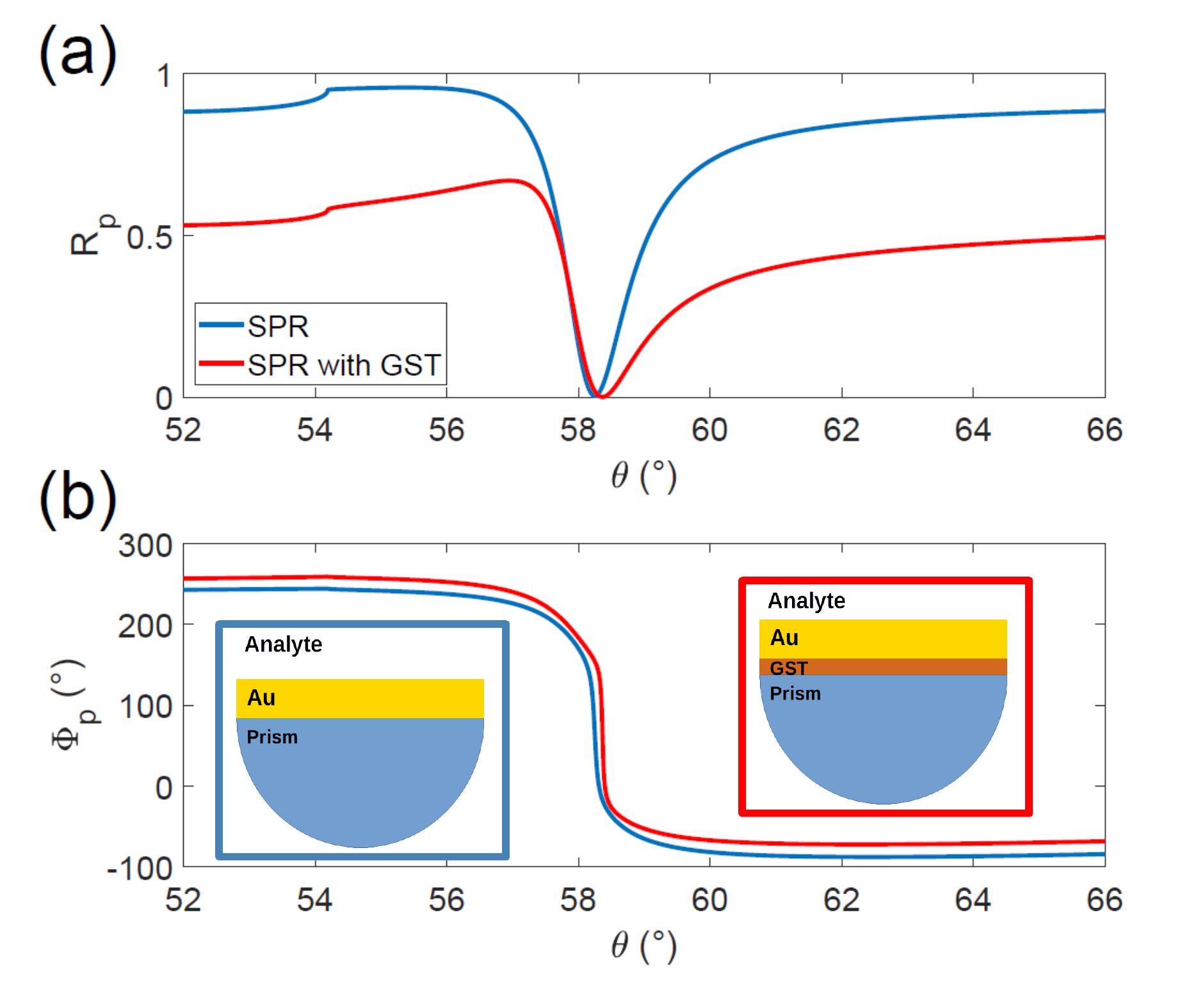}
\caption{Comparison of SPR and PCM-SPR devices: (a) Intensity and (b) phase of reflected light for a SPR device with 49.5nm (blue line) and PCM-SPR device with 5nm of GST on top of a 49.5nm gold layer (red line). The crystallization fraction of GST is $\tau=33 \%$.}
\label{fig_SPRwithGST_Inset}
\end{figure}

\begin{figure*}[htb]
\centering
\includegraphics[width=16cm]{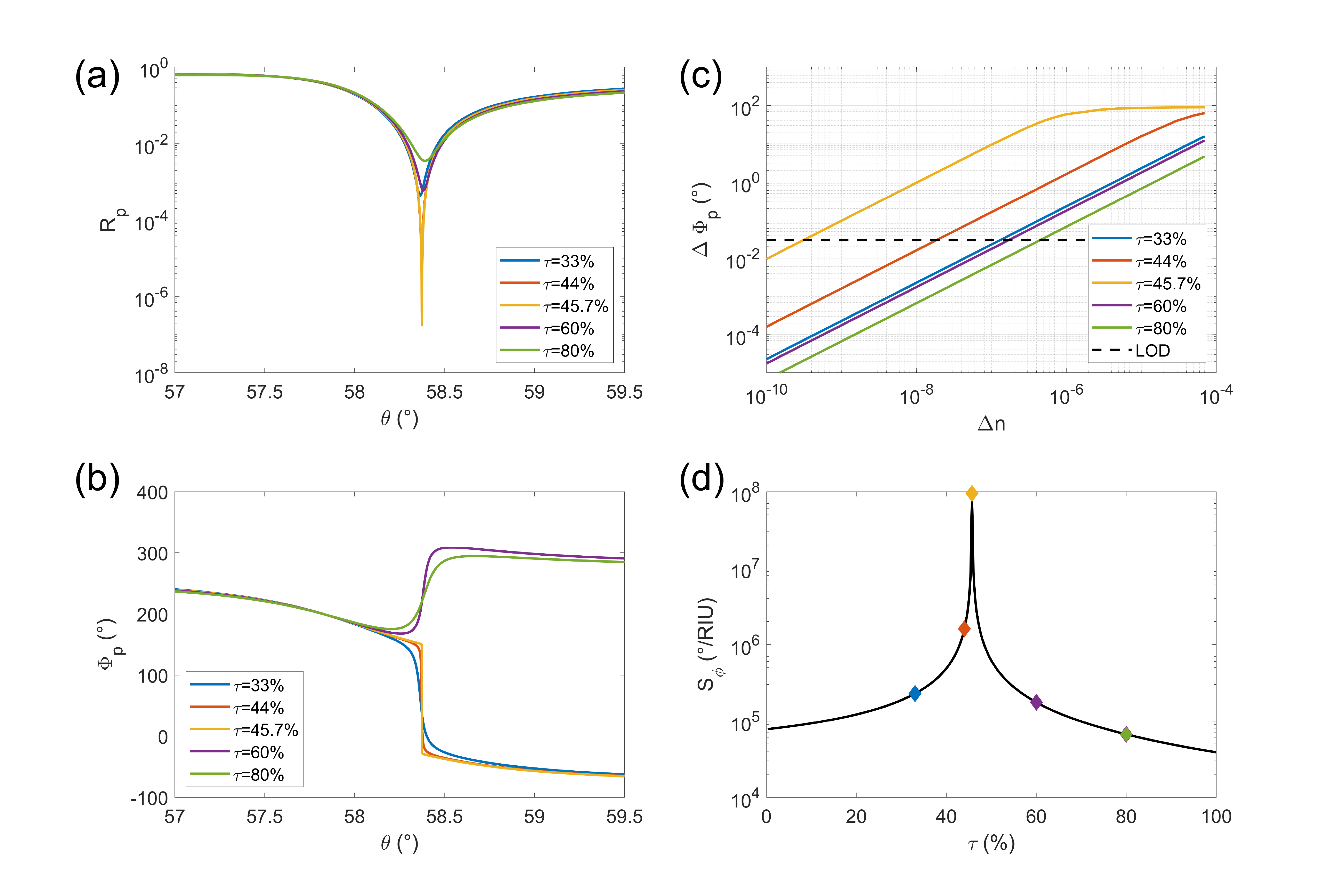}
\caption{Sensitivity of the PCM-SPR sensor : (a) Intensity and (b) phase of SPR reflected light for the PCM-SPR device with 5nm of GST on top of a 49.5nm gold layer for 5 different crystallization fraction. (c) Phase variation as a function of the refractive index variation  of the analyte for 5 different crystallization fraction. The dashed curve is the limit of measurable phase value  $3\sigma_{\phi}=0.03^{\circ}$ (d) Phase sensitivity as a function of the GST crystallization fraction $\tau$. The diamond markers are the sensitivity of the 5 curves in c)}
\label{fig_SPRwithGST}
\end{figure*}

We first analyze the impact of implementing a GST thin film below a metallic surface and the resulting effect on the SPR behavior. As exemplified in Fig. \ref{fig_SPRwithGST_Inset}, we show it is possible to excite a surface plasmon resonance in a 49.5 nm gold layer on top of a 5 nm GST layer at a wavelength of 750 nm. Even though the resonance is broadened, the deep absorption, characterized by a zero in reflectivity, is preserved. In other words the introduction of GST below the gold layer does not disturb the surface plasmon resonance.

As explained previously, upon crystallisation of GST, the absorption coefficient $k$ largely increases at the selected wavelength of 750 nm  while the refractive index does not significantly change (fig. \ref{fig_GSTdispersion}c)). Such a material can therefore be exploited as a dynamically tunable absorptive layer. By adding a thin GST layer between the gold and coupling medium, we are still able to  excite surface plasmons and to actively tune the absorption of the gold-GST system until perfectly equalizing the absorption and radiative leakages. The GST layer can therefore be used as a trimming layer to systematically reconfigure the system as a perfect absorber. In the following, we study the evolution of the critical coupling versus the crystallization of GST.  

In Fig. \ref{fig_SPRwithGST}(a) and (b) the intensity and phase of GST-assisted SPR are shown for five values of $\tau$. The minimum of reflectivity progressively decreases until reaching a value lower than $10^{-6}$ at the resonance angle when the crystallization fraction $\tau$ equals 45.7\% (Fig. \ref{fig_SPRwithGST}(a)). This reflectivity minimum also corresponds to an abrupt phase jump, as shown in Fig. \ref{fig_SPRwithGST}(b). 
Note that these results are qualitatively equivalent to those in Fig. \ref{fig_SPRsensitivity}, but the adjusting parameter is now the crystallized fraction $\tau$ of GST instead of the gold thickness. 

The calculated phase variation $\Delta{\phi}$ when the refractive index of the analyte is changing is displayed in Fig. \ref{fig_SPRwithGST}(c) for the five crystallisation fractions shown in Fig. \ref{fig_SPRwithGST}(a) and (b). The phase shift is linear in a wide range of refractive index variation and changes drastically with the crystallized fraction $\tau$ of GST. We find an optimal value of LOD as low as $3 \times 10^{-10}$ RIU at a crystallization fraction $\tau$=47.5\%, resulting in the lowest reflection. From there, we calculate the sensitivity $S_{\phi}$ and display it as a function of $\tau$ in fig. \ref{fig_SPRwithGST}(d). Strikingly, we demonstrate that the sensitivity of the system can be finely adjusted by three orders of magnitude via the crystallisation of GST and with predicted values of $S_{\phi}$ as high as $10^{8  \circ}/RIU$. Once this optimized value has been reached via the partial crystallization of GST, the system will stay in this configuration thanks to the non-volatile character of the crystallization of GST.


\subsection{Discussion}
We propose in this work a new approach to achieve ultimate phase sensitivities in surface plasmon sensors, up to a theoretical LOD of $3\times10^{-10}$ RIU. In the following, we highlight the usefulness of our approach by calculating typical amounts of chemical species that could be detected with such an ultimate sensitivity. To do so, we have selected four other works and extrapolated their experimental SPR measurements with our simulated LOD for PCM-SPR system.
Giorgini et al. \cite{Giorgini2018} are able to detect a covalent binding of streptavidin (around 50 kDa) with a surface coverage of $90~fg/mm^2$ thanks to a LOD of $10^{-8} RIU$. Since our calculated LOD is 33 times lower, the minimum detectable streptavidin surface coverage we should detect is $2.7~fg/mm^2$. 
For a gold standard in biosensors, Chou et al. \cite{Chou2006}, measured a concentration of 50pg/ml of mouse IgG (around 150 kD molecular weight) in its binding  with immobilized anti-mouse IgG with an SNR of 24 and a LOD of $3.5\times10^{-7}$ RIU . Since our LOD is three orders of magnitude better, the theoretical limit of detectable concentration of mouse IgG  would be $0.35 fg/ml$ i.e  2.3 attoM. For cancer diagnostic, in drug screening based on quantitative detection of protein cell death marker cytochrome c (around 12 kD molecular weight), our proposed system would detect a concentration of cytochrome c of 0.75 pM  i.e 60 times better than in \cite{Ng2013,Loo2014}. For volatile organic compounds (VOC) detection, we would obtain an LOD of 0.1 ppb (party per billion)\cite{Brenet2018}. 

In order to reach the very high phase sensitivity presented previously, the crystallization fraction $\tau$ of GST has to be progressively adjusted with a fine sampling. Such a fine tuning of partial crystallization was already demonstrated by different groups \cite{wang2014,cheng2018device,wang2021scheme}. In a recent work (\cite{Cueff2020}) we have demonstrated that critical coupling in a system comprising GST thin films, a dielectric layer and a gold layer, could be dynamically achieved by thermally crystallizing GST via a sampling scheme including more than a hundred partial crystallization states. 
\\
All the designs have been done at a wavelength of 750 nm, however any wavelength in the visible and near-infrared range could be used and would lead to similar results since in the visible and near-infrared range the phase change from amorphous to crystalline phase is always coming with a drastic change of the absorption coefficient $k$ allowing to tune the absorption (see supplement1 Fig. S1).

\begin{figure}[htbp]
\centering
\includegraphics[width=8 cm]{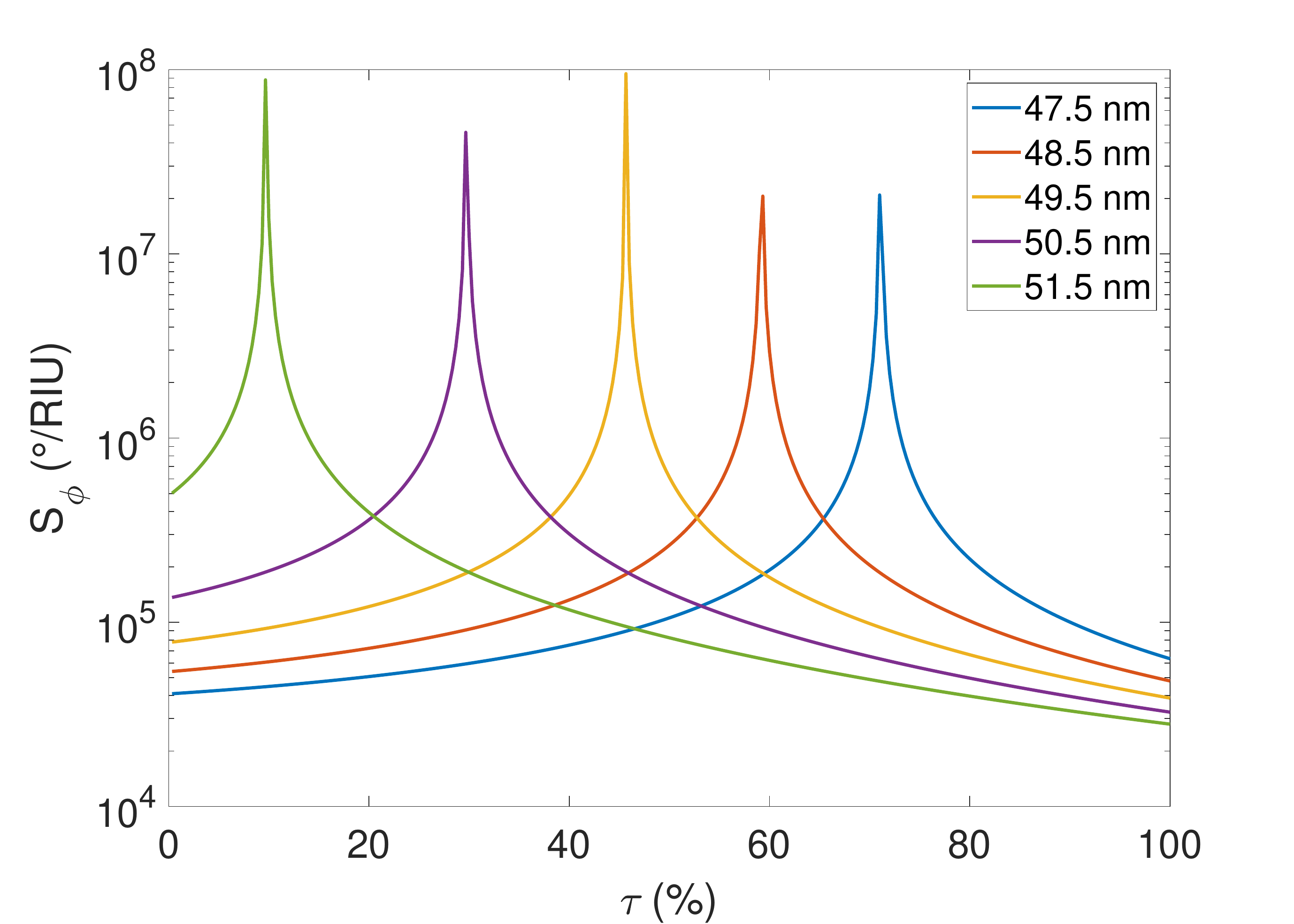}
\caption{Phase sensitivity as a function of the GST crystallization fraction $\tau$ for different gold thicknesses and a 5-nm-thick GST layer.}
\label{fig_SensitivityvsAuthickness}
\end{figure}

Note that the thickness values chosen for gold and GST in Fig. \ref{fig_SPRwithGST} are just an example of a functional system to reach an ultimate sensitivity. However, many different GST/Au thickness values can be chosen to reach similarly high sensitivity. If we keep a 5-nm-thick GST layer but use various thicknesses of gold, e.g. between 47.5 nm and 51.5 nm, we can see in Fig. \ref{fig_SensitivityvsAuthickness} that very high sensitivities can occur for each system, but at different crystallization fractions $\tau$ ranging from $9.7\%$ to $71\% $. Fig. \ref{fig_SensitivityvsAuthickness} shows that this PCM-assisted reconfigurable critical coupling is a very robust method to tune and achieve ultimate sensitivities thanks to the multilevel phase properties of PCMs. 
In other word we do not need precision on gold thickness deposition to achieve ultimate sensitivity thanks to the tunability of GST.  Similarly, other thicknesses of GST can yield ultimate sensitivities. 

This highlights two very important aspects of our proposed hybrid PCM-SPR approach: i) it enables the design of sensors with ultimate phase sensitivities; ii) it creates extremely robust and fabrication-tolerant systems in which any deviations from the targeted thicknesses can be reconfigured to reach the ideal conditions.
Indeed, high precision in the GST and gold layer thicknesses is no longer required since the critical coupling can be adjusted in real time by tuning the crystallization fraction while monitoring the minimal reflectivity in spectral measurements. Once the system is optimally reconfigured, as the tuning of GST is non-volatile, the device should be stable for years. \\

\section{Conclusion}
We have theoretically and numerically revisited the SPR in phase interrogation. To the best of our knowledge we establish for the first time in SPR a theoretical and quantitative relation between phase sensitivity, reflectivity at resonance and sensitivity in wavelength interrogation. From there, we explicitly show the superiority of phase interrogation SPR in detection sensitivity compared to intensity interrogation and we highlight the role of critical coupling in phase SPR sensors. This analysis provides a rational clarification on the large fluctuations reported in the phase sensitivities of SPR sensors measured by different groups, which result from non-optimal thicknesses of the metal layer. However, we also reveal that the ultimate phase sensitivity is hardly reachable since subnanometric precision would be required for the metal thickness, a condition that is ruled out by the roughness of the metal. We propose a practical solution to overcome this drawback by coupling the surface plasmon resonance with a layer of GST that can be reconfigured in a multilevel phase state. With that system, on can expect theoretical phase sensitivities with a limit of detection as low as $1\times10^{-10}$ RIU, i.e. one order of magnitude better that the previous reports in SPR phase measurements \cite{Li2008}. The presented methodology of assisting and tuning the critical coupling with nonvolatile reconfigurable multilevel PCMs can be applied to design new highly sensitive sensors in nanophotonics with phase singularities and topological darkness and will contribute towards the detection of trace amounts of analytes with low molecular weight.

\section*{Funding}
We acknowledge funding from the French National
Research Agency (ANR) under the project MetaOnDemand
(ANR-20-CE24-0013).

\medskip

\section*{Supplemental document}
See Supplement 1 for supporting content.

\bigskip

\appendix
\section{Complex reflection coefficient with coupled mode theory}
Using Coupled Mode Theory (CMT), the complex reflection coefficient of the surface plasmon resonance can be expressed: 
\begin{equation}
r(\omega,\theta)=\frac{\left(\frac{1}{\tau_{rad}}-\frac{1}{\tau_i}\right)-j\left(\omega-\omega_{sp}(\theta)\right)}{\left(\frac{1}{\tau_{rad}}+\frac{1}{\tau_i}\right)+j\left(\omega-\omega_{sp}(\theta)\right)}
\label{Equ_TMC_r}
\end{equation}
Where $\omega_{sp}(\theta)$ is the resonance pulsation for an incident angle $\theta$ and $\frac{1}{\tau_{i}}$ and $\frac{1}{\tau_{rad}}$ are the internal damping and external leakage respectively.\\
As $\omega=\frac{c}{n_0}\frac{2\pi}{\lambda}$ and $\omega_{sp}(\theta)=\frac{c}{n_{eff}^{sp}}\frac{2\pi}{\lambda}\sin(\theta)$, \ref{Equ_TMC_r} can be written:
\begin{equation}
r(\omega,\theta)=\frac{\frac{n_0n_{eff}^{sp}}{c}\frac{\lambda}{2\pi}\left(\frac{1}{\tau_{rad}}-\frac{1}{\tau_i}\right)-j\left(n_{eff}^{sp}-n_o\sin(\theta)\right)}{\frac{n_0n_{eff}^{sp}}{c}\frac{\lambda}{2\pi}\left(\frac{1}{\tau_{rad}}+\frac{1}{\tau_i}\right)+j\left(n_{eff}^{sp}-n_o\sin(\theta)\right)}
\label{Equ_TMC_r_2}
\end{equation}
And gives equation (3) with:
\begin{equation}
\gamma_{rad,i}=\frac{n_on_{eff}^{sp}}{c}\frac{\lambda}{2\pi}\frac{1}{\tau_{rad,i}}
\end{equation}

\section{Phase Sensitivity}
From \ref{Equ_TMC_r} we get the reflection phase and the minimum reflectivity:

\begin{align}
\phi_r&=-atan\left(\frac{\omega-\omega_{sp}(\theta)}{\frac{1}{\tau_{rad}}-\frac{1}{\tau_i}}\right)-atan\left(\frac{\omega-\omega_{sp}(\theta)}{\frac{1}{\tau_{rad}}+\frac{1}{\tau_i}}\right)\\
\phi_r&\sim-\frac{\omega-\omega_{sp}(\theta)}{\frac{1}{\tau_{rad}}-\frac{1}{\tau_i}}-\frac{\omega-\omega_{sp}(\theta)}{\frac{1}{\tau_{rad}}+\frac{1}{\tau_i}}\\
r_{min}&=r(\omega_{sp},\theta)=\frac{\frac{1}{\tau_{rad}}-\frac{1}{\tau_i}}{\frac{1}{\tau_{rad}}+\frac{1}{\tau_i}}
\end{align}

Which leads to:
\begin{equation}
\phi_r=-2\frac{\omega-\omega_{sp}(\theta)}{\omega_{sp}(\theta)}\left(\frac{1}{r_{min}}+1\right)Q
\end{equation}

Where $Q$ is the quality factor of the resonance given by:
\begin{equation}
Q=\frac{\omega_{sp}(\theta)}{2}\frac{1}{\frac{1}{\tau_{rad}}+\frac{1}{\tau_i}}
\end{equation}

Near the critical coupling $r_{min}\ll1$ and we get:

\begin{equation}
\phi_r=-2\frac{\omega-\omega_{sp}(\theta)}{\omega_{sp}(\theta)}\frac{Q}{r_{min}}
\label{Equ_TMC_phase}
\end{equation}

The measured experimental sensitivity  $S_{\phi}$ is related to the phase variation when refractive index of analyte is changing. The definition is :

\begin{equation}
S_{\phi}=\frac{d\phi}{dn}
\end{equation}

When the refractive index of the analyte changes, the pulsation of resonance $\omega_{0}$ shifts. Following equation \ref{Equ_TMC_phase}, the phase derivation versus the pulsation of resonance is :
\begin{equation}
\frac{d\phi}{d\omega_{sp}(\theta)}=-\frac{2\omega}{\omega_{sp}^2(\theta)}\frac{Q}{r_{min}}
\end{equation}
At the resonance ${\omega}$=${\omega_{sp}}$ hence:
\begin{equation}
\frac{d\phi}{d\omega_{sp}}=-\frac{2}{\omega_{sp}(\theta)}\frac{Q}{r_{min}}
\label{DerivePhasevsEnergie}
\end{equation}

The units for the phase sensitivity is $^{\circ}/RIU$, with RIU the refractive index units.
We can write the relations between sensitivity  $S_{\phi}$ and the derivative of the phase $\phi$ if we make appear the wavelength shift of the resonance $\lambda_{sp}$ in the equation.
\begin{equation}
S_{\phi}=\frac{d\phi}{dn}=\frac{d\phi}{d\lambda_{sp}}\frac{d\lambda_{sp}}{dn}=\frac{d\phi}{d\omega_{sp}}\frac{d\omega_{sp}}{d\lambda_{sp}}\frac{d\lambda_{sp}}{dn}
\label{SensibilitevsLongueurdOnde}
\end{equation}

 $\frac{d\omega_{sp}}{d\lambda_{sp}}$ can be connected to speed of light $c$ in the detector space.

\begin{equation}
\frac{d\omega_{sp}}{d\lambda_{sp}}=-\frac{2\pi}{\lambda_{sp}^2}\frac{c}{n_{o}}
\label{RelDerivDispersion}
\end{equation}
Equations \ref{DerivePhasevsEnergie} and \ref{RelDerivDispersion} in equation \ref{SensibilitevsLongueurdOnde} leads to:

\begin{equation}
S_{\phi}=\frac{2}{\lambda_{sp}}\frac{Q}{r_{min}}\frac{d\lambda_{sp}}{dn}
\label{SensibilitevsLambda}
\end{equation}

$\frac{d\lambda_{sp}}{dn}$ is the resonance shift versus refractive index of the analyte and is usually measured in spectroscopic measurements. In others words it is the transducer wavelength sensitivity  that we will name $S_{\lambda}$ and hence we establish a relation between phase sensitivity $S_{\phi}$ and wavelength sensitivity $S_{\lambda}$:
\begin{equation}
S_{\phi}=\frac{2}{\lambda_{sp}}\frac{Q}{r_{min}}S_\lambda
\label{SensibilitevsLambda}
\end{equation}


\bibliography{Biblio_SPR_PCM4}

\end{document}